\documentclass[nofootinbib, showpacs, showkeys,
amsmath, amssymb]{revtex4}

\begin{document}

\title{ Random versus holographic fluctuations of the background metric. II. \\ Note on the dark energies arising due to
microstructure of space-time }

\author{ Michael~Maziashvili}
\email{maziashvili@gmail.com} \affiliation{Andronikashvili
Institute of Physics, 6 Tamarashvili St., Tbilisi 0177, Georgia \\ Center for Elementary Particle Physics, ITP, Ilia State University, 3\,$-$\,5 Cholokashvili Ave., Tbilisi 0162, Georgia  }

\begin{abstract}

Over the last few years a certain class of dark-energy models decaying inversely
proportional to the square of the horizon distance emerged on the basis either of Heisenberg
uncertainty relations or of the uncertainty relation between the four-volume and the cosmological
constant. The very nature of these dark energies is understood to be the same, namely it is
the energy of background space/metric ﬂuctuations. Putting together these uncertainty relations
one ﬁnds that the model of random ﬂuctuations of the background metric is favored over the
holographic one.

\end{abstract}





\maketitle

Space-time uncertainty is common for all approaches to quantum
gravity. Quantum gravity strongly indicates the finite resolution
of space-time; it will suffice to mention a few well known
examples: space-time uncertainty relations in string theory
\cite{String1, String2}; noncommutative space-time approach
\cite{noncommutative}; loop quantum gravity \cite{Loop}; or
space-time uncertainty relations coming from a simple {\tt
Gedankenexperiments} of space-time measurement
\cite{Gedankenexperiments}. Well known entropy bounds emerging via
the merging of quantum theory and general relativity also imply
finite space-time resolution \cite{entropybounds}. The combination
of quantum theory and general relativity in one or another way
manifests that the conventional notion of distance breaks down at least at the Planck scale $l_P \simeq 10^{-33}$\,cm
\cite{minimumlength}. Indeed, the finite space-time resolution can
readily be shown in simple physical terms. (In what follows we
will assume the system of units $\hbar = c = 1$). Namely, posing the
question as to what maximal precision one can mark a point in space
by placing there a test particle, one notices that in the
framework of quantum field theory the quantum takes up at least
the volume, $\delta x^3$, defined by its Compton wavelength
$\delta x \gtrsim 1/m$. Not to collapse into a black hole, general
relativity insists the quantum on taking up a finite amount of
room defined by its gravitational radius $\delta x \gtrsim
l_P^2m$. Combining together both quantum mechanical and general
relativistic requirements one finds
\begin{equation}\label{abslimit} \delta x \,\gtrsim\,
\mbox{max}(m^{-1},~l_P^2m)~.\end{equation} From this equation one
sees that a quantum occupies at least the volume $ \sim l_P^3 $.
Since our understanding of time is tightly related to the periodic
motion along some length scale, this result implies in general the
impossibility of space\,-\,time distance measurement to a better
accuracy than $\sim l_P$. Therefore, the point in space\,-\,time
can not be marked (measured) to a better accuracy than $ \sim
l_P^4$. It is tantamount to say that the space\,-\,time point
undergoes fluctuations of the order of $\sim l_P^4$. Over the
space\,-\,time region $l^4$ these local fluctuations add up in
this or another way that results in four volume fluctuation of
$l^4$. In view of the fact of how the local fluctuations of
space\,-\,time add up over the macroscopic scale ($l \gg l_P$),
different scenarios come into play. Most interesting in quantum
gravity are random and holographic fluctuations. If the local
fluctuations, $\sim l_P$, are of random nature then over the
length scale $l$ they add up as $\delta l = (l/l_p)^{1/2}l_P$. In
the holographic case, the local fluctuations, $\sim l_P$, add up
over the length scale $l$ to $\delta l =
(l/l_P)^{1/3}l_P$.\footnote{It is curious to notice that combining
quantum mechanics with general relativity, the relation $\delta l
\gtrsim l_P^{2/3}l^{1/3}$ as an intrinsic imprecision in measuring
of length scale $l$ (for the Minkowskian background space)  was
obtained by K\'arolyh\'azy in 1966 \cite{Gedankenexperiments}.}
Albeit
locally (that is, at each point) the space\,-\,time undergoes
fluctuations of the order of $\sim l_P^4$, for the fluctuations add up
over the length scale $l$ to $\delta l(l)$, the region $l^4$
effectively looks as being made of cells $\delta l(l)^4$. Such a cell represents a minimal detectable unit of space-time over a region $l^4$ (in other words, it determines the maximum precision by which the volume $l^4$ can be known); that
immediately prompts the rate of volume fluctuations $\delta V = \delta l(l)^4$. This is the quantity that enteres the uncertainty relation between the four volume and the cosmological constant.

Treating $\Lambda m_P^2/8\pi$ and four volume as conjugate to each
other in the same sense as energy and time are conjugate in
quantum mechanics, one can invoke the uncertainty principle to
predict \cite{I}
\begin{equation}\label{volccrelation}\rho_{\Lambda} \,\simeq\, {1 \over \delta
V}~,~~~~\mbox{where}~~~~\rho_{\Lambda} \,\equiv\, {\Lambda m_P^2
\over 8\pi}~.\end{equation} So that the $\rho_{\Lambda}$ in
Eq.(\ref{volccrelation}) takes the definite meaning to be the local
quantity
\begin{equation}\label{rhocc} \rho_{\Lambda} \,\simeq\, {1 \over
\delta l^4}~.\end{equation}  In the case of random fluctuations,
$\delta l \simeq (l_Pl)^{1/2}$, from Eq.(\ref{rhocc}) one gets
pretty good value of dark-energy density for the present epoch $l
\simeq H_0^{-1}$ (where the present value of Hubble parameter $H_0
\simeq 10^{-61}m_P$) \cite{darkI}
\begin{equation}\label{inversesquare} \rho_{\Lambda} \,\simeq\, {1
\over l_P^2\,l^2  }~. \end{equation}

However, we expect another dark energy to be also present simply
because of time-energy uncertainty relation. Because of
fluctuations, the length scale $l$ can be known with a maximum
precision $\delta l$ determining thereby the minimal detectable
cell $\delta l^3 $ over the spatial region $l^3$. Such a cell
represents the minimal detectable unit of space-time over a given
length scale and if it has a finite age $t$, its existence due to
time-energy uncertainty relation can not be justified with energy
smaller then $\sim t^{-1}$. Hence, if the age of the background
space-time is $t$ then because of time energy uncertainty relation
the existence of minimal cell $\delta l^3$ over the region $l^3$
can not be justified with energy smaller than

\[  E_{\delta l^3} \,\gtrsim\, t^{-1}~.
\] Hence, time-energy uncertainty relation tells us
that there should be another kind of dark energy \cite{II}
\begin{equation}\label{cellenergy} \rho \,\simeq\, {E_{\delta l^3}\over \delta l^3} \,=\, {
1 \over t\delta l^3 }~.\end{equation} For the K\'arolyh\'azy
uncertainty relation, $\delta l \simeq l_P^{2/3}l^{1/3}$, it gives

\[ \rho \,\simeq\, {1 \over l_P^2\,l\, t }~, \] which, by taking into account that during radiation and matter
dominated epochs $t \simeq l$, is nothing else but Eq.(\ref{inversesquare}) \cite{II, darkII}.

However, from Eqs.(\ref{rhocc},\,\ref{cellenergy}) one sees that
$\rho \ll \rho_{\Lambda} $ as long as $t \gg \delta l$. For
holographic fluctuations
\[ \rho_{\Lambda} \,\simeq \, {1 \over
\delta l^4} \,\simeq \,{1 \over l_P^{8/3}l^{4/3}}~,\]
$\rho_{\Lambda}$ becomes unacceptably large $\rho_{\Lambda}(l
\simeq 10^{61}l_P) \simeq 10^{-81}m_P^4$. On the other hand, for
random fluctuations,
\[  \rho \,\simeq\,  {
1 \over t\delta l^3 } \,\simeq\,  { 1 \over  l_P^{3/2}l^{5/2}} ~,\]
$\rho $ is negligibly small in comparison with the observed value
of dark energy density $\rho(l \simeq 10^{61}l_P) \simeq
10^{-152}m_P^4$.

The above discussion demonstrates that putting the four
volume\,-\,cosmological constant uncertainty relation on the equal
footing with the Heisenberg uncertainty relations \cite{I}, the
model of random fluctuations appears to be favored over the
holographic one. Curiously enough, the uncertainty relation
$\delta l = (l_Pl)^{1/2}$ is favored over the K\'arolyh\'azy
uncertainty relation, $\delta l = l_P^{2/3}l^{1/3}$, also in the
framework of {\tt Gedankenexperiments} for space-time measurement as well as in the context of quantum-gravitational running of space-time dimension
\cite{mazia}. It is important to keep in mind that the nature of
both dark energies considered above is the same, namely it is the
energy of background space/metric fluctuations.

\vspace{0.2cm}

The work was supported in part by the \emph{CRDF/GRDF} grant.

\end{document}